# COCCIA LAB

*To discover the causes of social, economic and technological change*

# Effects of the institutional change based on democratization on origin and diffusion of technological innovation

Mario COCCIA

CNR -- National Research Council of Italy
&
Yale University

# Effects of the institutional change based on democratization on origin and diffusion of technological innovation

*Mario Coccia[1]*

CNR -- National Research Council of Italy & Yale University

CNR -- National Research Council of Italy
Collegio Carlo Alberto, Via Real Collegio, 30-10024 Moncalieri (Torino, Italy)
Yale School of Medicine
310 Cedar Street, Lauder Hall, New Haven, CT 06510, USA
*E*-mail: mario.coccia@cnr.it

Mario Coccia 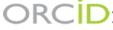: http://orcid.org/0000-0003-1957-6731

## Abstract

Political systems shape institutions and govern institutional change supporting economic performance, production and diffusion of technological innovation. This study shows, using global data of countries, that institutional change, based on a progressive democratization of countries, is a driving force of inventions, adoption and diffusion of innovations in society. The relation between technological innovation and level of democracy can be explained with following factors: higher economic freedom in society, effective regulation, higher economic and political stability, higher investments in R&D and higher education, good economic governance and higher level of education system for training high-skilled human resources. Overall, then, the positive associations between institutional change, based on a process of democratization, and paths of technological innovation can sustain best practices of political economy for the development of economies in the presence of globalization and geographical expansion of markets

**Citation:** Coccia M., 2020. Effects of the institutional change based on democratization on origin and diffusion of technological innovation, *Working Paper CocciaLab n. 44/2020, CNR -- National Research Council of Italy.*

[1] Acknowledgement. I gratefully acknowledge financial support from National Research Council of Italy–Direzione Generale Relazioni Internazionali for funding this research project developed at Yale University in 2019 (grant-cnr n. 62489-2018). The author declares that he has no relevant or material financial interests that relate to the research discussed in this paper.

# GOALS OF THE INVESTIGATION

Social scholars argue that the development of human societies over the long term is due to technological change or institutions that enable the definition and defense of formal property rights (Auerswald and Stefanotti, 2013; Coccia, 2010, 2019, 2019c). However, the interaction between these two concepts is hardly known. Chlebna and Simmie (2018) claim that while there is agreement among scholars on the importance of institutions with respect to economic and industrial development, there remains little analysis on *how* and *why* institutions interact with technological change on which industrial development of advanced and emerging economies is based. The main aim of this chapter is to explain, whenever possible, the relation between institutional change, based on a process of democratization, and the patterns of technological innovation across countries. A theoretical background of the concepts of institutions and institutional change is useful to understand and clarify this vital relation that can explain the paths of development in society.

Institutional theory explains both individual and organizational actions. A main research field of institutional theory is the analysis of *how* institutions change over time (Campbell, 2004; Dacin et al., 2002; Di Maggio et al., 1991; Williamson, 2000). First of all, a debate revolves around how to conceptualize *institutions* and *institutional change* (Roland, 2004). The literature suggests different definitions of institution, which affect the perspective to study institutional change in society (cf., Alston, 1996; Coccia, 2019; Kingston and Caballero, 2009; Hodgson, 2006; Milgrom et al., 1990). Veblen (1899, p. 190) argues that institutions are: "prevalent habits of thought with respect to particular relations and particular functions of the individual and of the community" (cf., Brette, 2003). Hayek (1973) considers institutions based on shared expectations in society, rather than rules. North (1990, p. 3; 2005) states that institutions: "are the rules of the game in a society, or more formally, are the humanly devised constraints that shape human interaction …. reduce uncertainty by providing a structure to everyday life". Auerswald and Stefanotti (2013, p. 113) state that institutions in general, and property rights in particular, are crucial to the functioning of credit markets that in turn are a key to economy-wide growth (cf., Coase, 1960; Demsetz, 1967). In general, institutions are based on formal rules (such as laws and constitutions) and informal constraints (such as, conventions and norms). Instead, Aoki (2001, 2007) defines institutions as stable and shared systems of beliefs about the expected behavior of the members of a society in various contingencies. Greif (2006, p. 30) adopts a broad definition of institution considering: "a system of rules, beliefs, norms and organizations that together generate a regularity of (social) behavior". In brief, North (1990) sees institutions as rules, whereas Aoki (2007, p. 6) views institutions as "selfsustaining, salient patterns of social interactions" that give rise to "common knowledge among the players regarding a particular equilibrium path of the game".

The literature also proposes different theoretical frameworks of institutional change (cf., Coccia, 2019; Kingston and Caballero, 2009). In North's approach, institutional change is an accumulation of incremental changes rather than occasional, radical changes. Libecap (1989) claims that institutional change is a path-dependent process in which institutions are a function of current technologies, but also of previous technologies and institutions. The institutional change is also a path-dependent process because individuals learn, organizations develop, and ideologies form in the context of formal and informal rules (Murat and Jared, 2017). Ostrom (2005) recognizes both exogenous causes of institutional change (e.g., technological change) and endogenous causes (e.g., the depletion of a resource over time). In particular, Ostrom (2005) distinguishes between "operational rules", which govern day-to-day interactions, "collective choice rules" (rules for choosing operational rules), and "constitutional rules" (rules for choosing collective-choice rules), whereas "meta constitutional rules" are for choosing constitutional rules (e.g., the "rules" by which a civil war is fought). Moreover, each individual calculates the expected costs and benefits of a given institutional change and, if a "minimum coalition" necessary to effect that change agrees to it, an institutional change can occur. Therefore, Libecap (1989) and Ostrom (2005) argue that an institutional change depends on higher-level rules and on how decision makers perceive the likely effects of a change in rules. Scholars also analyze institutional change as an evolutionary process (cf., Kingston and Caballero, 2009; Coccia, 2018, 2018c, 2019). Theories of evolutionary institutional change suggest that institutional change is due to human actions, such as learning, imitation, etc. The difference between evolutionary and designed-based theories of institutional change lies in the role of selection processes determining which rules emerge and adapt in socioeconomic environments (Coccia, 2019; cf., Coccia, 2017e). In particular, evolutionary theories do not consider a central mechanism (e.g., legislation) that affects interactions of people in society.

The interaction between institutional change and patterns of technological innovation has been analyzed with different perspectives (Coccia, 2019). Ayres (1944, p. 187) considers exogenous technological progress as the main driver of institutional change: "technological development forces change upon the institutional structure by changing the material setting in which it operates". Nelson (2005, p. 169) sees changes in physical technology as a source of institutional change. In general, technological evolution can be a determinant of institutional change in society (Coccia, 2018a; Coccia, 2019a, b; Coccia and Watts, 2020; Perez, 2004), though the relationship can be bi-directional, with interrelationships between technological change and institutional change (Coccia, 2010, 2014, 2014a, 2018, 2019, 2019a, b, c, d, e). In particular, institutions can affect technology generating an interaction, so that "it probably is useful to think of physical and social technologies as coevolving" (Nelson, 2005; cf., Coccia, 2010, 2014, 2018b, 2019, 2019a, 2019b, 2016c; Coccia and Watts, 2020).
Overall, then, economists and policymakers have increasingly recognized the role played by institutions and institutional change in the process of economic and technological development (Coccia, 2019). This contribution now moves on to discuss the relationships between institutional change, based on a process of democratization, and innovative outputs across countries, trying, as far as possible, to clarify these topics that are important, very important for supporting the economic growth of countries.

4 | P a g e
Coccia M. (2020) *Effects of the institutional change based on democratization on origin and diffusion of technological innovation*
*CocciaLab Working Paper 2020 – No. 44/2020*

# THEORETICAL FRAMEWORK

Literature shows different perspectives to investigate the role of institutions for technological change (Kingston and Caballero, 2009). Nelson (1993) considers institutions as the legislation and organization of education and training that differ at national level, and therefore form the basis of distinctive national systems of innovation. Edquist and Johnson (1997) define institutions as behavioral patterns such as routines, norms, shared expectations and morals. Lundvall and Maskell (2000) argue that institutions develop from and co-evolve with solving specific problems through processes of interactive learning (cf., Bathelt and Glückler, 2014; Coccia, 2016). Chlebna and Simmie (2018) observe that technical change requires complementary institutional change and that new technologies may not be supported by existing institutional arrangements (Freeman and Perez, 2008; Nelson, 1998). As a result, for major innovation to succeed "institutional and regulatory changes must take place" (Rip and Kemp, 1998, p. 364). North (1990) argues that the concept of path dependence can be applied to both technological and institutional change. In fact, Setterfield (1993, p. 761) also suggests that institutions can evolve with path-dependent phenomena. In general, institutions and institutional change play a significant role among the various forces of economies underlying the development of technological trajectories. Chlebna and Simmie (2018, p. 973) argue that some agents possess or develop the capacity to stimulate institutional change. In this context, Garud et al. (2007) identify the *institutional entrepreneurs* that have an interest in particular institutional arrangements and leverage resources to create new institutions or to transform existing ones. Socioeconomic movements can also play a key role as collective agents of institutional change (Doblinger and Soppe, 2013; Vasi, 2011). Chlebna and Simmie (2018) state that institutions can co-evolve with the introduction of technological innovations for them to diffuse through the economy. Chlebna and Simmie (2018) also suggest that informal institutions, through their impact on the behaviors of agents, influence the degree to which they press for formal institutional arrangements to coevolve with technological developments. Simultaneously, the degree of openness of formal and organizational institutions impacts on the ability of agents to foster institutional co-evolution. Martin (2008) argues that technological change, as an inherently socio-cultural activity, deeply depends on institutional setting within which it takes place. Moreover, informal institutions provide more fertile and less rigid environments for the generation of new ideas than formal and organizational institutions. In particular, the norms and beliefs that constitute informal institutions influence behaviors and the willingness of individuals, such as entrepreneurs consider new ideas to support change. In short, institutions form an important filter for the perceptions of agents with respect to interactions between technological trajectories and their wider environment. As a matter of fact, path-dependent technological trajectories are intertwined with their institutional settings so new path creation is also influenced by historical institutional arrangements and their co-evolution with the introduction of new technologies. Hence, co-evolving parts can both enable and constrain each other through feedback that can be negative or positive (Garud and Karnøe, 2001). In this context, Perez (2004) states that the deployment of each technology system involves several interconnected processes of change and adaptation: 1) development of surrounding services (required infrastructure, specialized suppliers, distributors, maintenance services, etc.) 2) "cultural" adaptation to the logic of interconnected technologies involved (among engineers, managers, sales and service people, consumers, etc.); 3) setting up of institutional facilitators (rules and regulations, specialized training and education, etc.).

Overall, then, the literature in this field of research is vast but it has not clarified the role of institutions and institutional change in technological innovation, such that the interactions between institutional change, based on process of democratization of countries, and origin and diffusion of technologies are hardly known (cf., Chlebna and Simmie, 2018). In particular, the fundamental questions in economics of innovation and institutional theory are:

- *What is the relationship between innovation and institutional change?*
- *Does innovation depend upon institutional change of democratization in society?*

- *What are differences between levels of innovative and economic performance across countries in terms of institutional change based on higher and/or lower democratization process?*
- *Why do some societies have higher innovative outputs, fixed the level of institutional change and democratization?*
- *How does institutional change, driven by democratization, affect the origin of innovative outputs, adoption and diffusion of new technologies across countries?*

This contribution confronts these questions to explain, whenever possible, the relationship between socio-institutional factors and elements of technological change, which can provide results to support technological, economic and social change of nations. In particular, the purpose is to determine *if* and *how* institutional change, based on democratization, affects paths of technological development across countries; in fact, this relation has main implications for political economy of growth to support institutional and innovation policies of countries that fertilize the economic system and underpin the technological and economic development in society. Studies show that institutional structure and political system of countries can be – through law, social rules and education system – driving forces for technical change in society (Coccia, 2010, 2012, 2015, 2017a, b, c). In particular, a main relationship is between innovative outputs and level of institutional change directed to democratization of nations (Coccia, 2019). Democracy can be seen as a set of practices and principles that institutionalize and protect freedom (cf., Bobbio, 2005, 2006; Mosca, 1933; Pareto, 1946). Most scholars would agree that the fundamental features of a democracy include a government based on majority rule and the consent of governed, the existence of free and fair elections, the protection of minorities and respect for basic human rights (Norris, 2008). In fact, the Schumpeterian minimalist conception of democracy is a political system based on elections[2] (Schumpeter,

---

[2] "The democratic method is that institutional arrangement for arriving at political decisions in which individuals acquire the power to decide by means of a competitive struggle for the people's vote" (Schumpeter, 1942, p. 269).

1942). Przeworski et al. (2000) consider democracy as the political system in which key government offices are filled through contested elections. Democracy presupposes equality before the law, because of political pluralism, whereas democratization is a process of institutional change that improves laws and institutions for supporting the wellbeing of people and wealth of nations. Several researches have showed that democracy has been increasing over time. In particular, Modelski and Perry III (2002) consider democratization as a long-run process of social innovation that has taken 120 years to move from 10% to 50% across countries (roughly in year 2000), whereas 90% of institutional democratization will be achieved in the 2110s or thereabouts. As a matter of fact, democracy, by a Darwinian process of natural selection, seems to be the best political system that survives to social change, absorbs and supports economic and technological change. In addition, the proposition that wealthy society is usually also more democratic has a long lineage (Lipset Seymour, 1959). This hypothesis has been confirmed by Barro (1999), though the precise effect is sensitive to each time-period analyzed, to the selection of control variables specified in models, and to the measurement of both democracy and economic growth. Barro (1999, p. 160) points out that "increases in various measures of the standard of living forecast a gradual rise in democracy". Norris (2008) and other scholars argue that democratization comes together with economic growth (cf., Tavares and Wacziarg, 2001). Conversely, Persson and Tabellini (2003, 2007) claim that constitutional arrangements have the ability to influence economic policies and economic performance, and thus patterns of socio-economic development. Therefore, democracy may have effects on economic growth. Acemouglu et al. (2008) revisit the relationship between income per capita and democracy and argue that political and economic development paths are mainly interwoven. The economic debate has also examined how the institutional change of democratization can affect the patterns of technological innovation across countries. In particular, Coccia (2010) shows that new democratic laws in England and France, as well as the United States constitution of 1791, can be considered as the socio-economic background of institutions and institutional change for the origin and diffusion of the First and Second Industrial Revolution based on major technological innovations (e.g., steam engine, spinning jenny, etc.) that changed the socio-economic structure of European and North-American economies, generating an exceptional increase in employment, wealth and economic growth of nations (Figure 1).

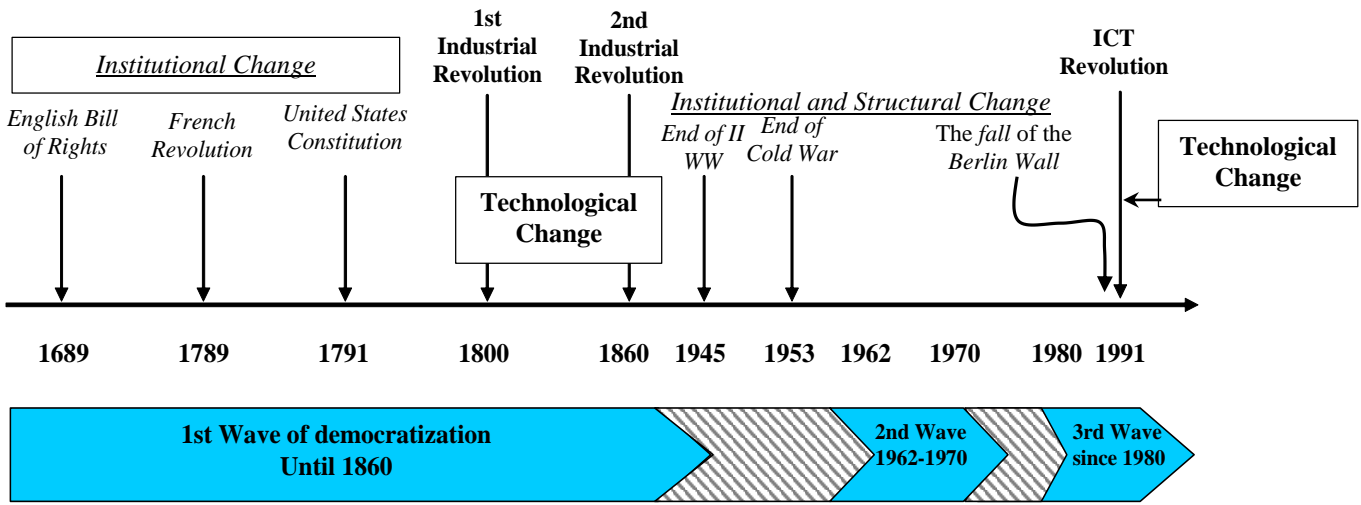

Figure 1 – Institutional change and new institutions, based on democratization, as preconditions to technological revolutions (adapted from Coccia, 2010)

As a matter of fact, the civil war in England (1688), the revolution of the American colonies (between 1775 and 1783) and the French revolution (1789–1799) generated a variety of social and political forces, new institutions and a fruitful institutional change that reduced social and cultural friction and led to the exploiting of path-breaking inventions, such as the steam engine supporting accelerated rates of employment and economic growth in Europe and North America (cf. also, Coccia, 2010, 2018c, 2019h). Mokyr (2002) argues that the scientific revolution and the Enlightenment movement in Europe (from 16$^{th}$ to 18$^{th}$ Centuries) helped expand the epistemic base of techniques in use and created the social conditions for technological and economic progress. In fact, the Industrial Revolution requires not just new knowledge and technology but also of appropriate institutions that sustain the ability of society to access this knowledge/technology, use it, improve it, and find new applications and combinations for it in society. Headrick (2000) claims that the age of industrial revolution, through a variety of technological and institutional innovations, created a new political and social climate that supported more democratic countries. Had the institutional feedback been negative as it had been before 1750, technological progress would have been on the whole short-lived (cf., Coccia, 2018b). Yet the feedback between institutions and technology was and is positive (Coccia, 2010). In particular, the years after 1815 were more and more subjugated by the free market liberal ideology, which provided incentives for scientific discoveries and entrepreneurship within more democratic countries. Moreover, new democracies emerging in the late 20th Century has renewed interest in the relationship between democracy and economic performance (Huntington,1991; Kurzman, 1998). In general, liberal democracy (with effective legal system and political competition) can support a good economic governance that will translate into improved social cohesion and economic performance of nations (Acemoglu, 2018; cf., Farazmand and Pinkowski, 2006; Farazmand, 2019). Kyriazis and Karayiannis (2011) suggest a new theoretical perspective on democracy as a system that facilitates changes, especially in the form of direct democracy. They stress the role of the initiator, i.e., anybody who has the right to introduce a new proposal. Decision makers here can choose strategies form this set, and under a continuous process of trial and error can reject wrong ones and retain correct ones (in the sense of welfare increasing strategies). Thus, society can gain knowledge and new efficient institutions emerge. Taverdi et al. (2019) show that the level of democracy affects the quality of governance and confirm that political freedom and civil rights influence the level of governance with a non-linear effect. In fact, governance quality is typically weaker in countries with intermediate levels of political freedom than in their less democratic counterparts, but once past the threshold level, greater political competition is associated with stronger governance. Countries, with a consolidated process of democratization, experience a much higher quality of governance that is the background for fruitful economic, technological and social change. Taverdi et al. (2019) also suggest that the effectiveness of governance increases with economic development and education (cf., Castelló-Climent, 2008). In short, higher economic and state freedom enhances governance. Nevertheless, large population, unequal distribution of income and natural resource abundance can reduce governance quality. Other studies by Kotschy and Sunde (2017) point out that excessively high levels of inequality erode institutional quality even in democracies, up to the point that democracies appear not to be able to implement good institutional environments if inequality is too high. To put it differently, as said, there is a non-linear relationship between different level of governance and democracy across countries. Policy implications are that effective and efficient democratic institutions to support a good quality governance, control corruption and generally allow the state to achieve its social and economic objectives in the long run. In short, effective institutions require a high level of transparency, participation and representation, which in turn strengthen the quality of governance. In addition, transition countries can overcome the problem of weak governance once the democratic consolidation has been achieved (cf., Lindseth, 2017; Aidt and Jensen, 2013; Bartlett, 1996).

Bedock et al. (2012) argue that institutional change of advanced and consolidated democracies can be due to legitimacy problems, socioeconomic issues, technological and social development, policy diffusion and globalization of economies.

This theoretical background, just described, supports the analyses and results of a study here on these topics.

# MATERIALS AND METHODS

*1.1  Data and their sources*

The sample under study here is 191 countries. Sources of data concerning the institutional change are from the OECD (2013), the World Bank (2008), the Worldwide Governance Indicators (2019) and Norris (2008a). Data of technological innovation outputs are taken from World Bank (2009) and Norris (2008a).

*1.2  Measures*

− Institutions and institutional change

This contribution measures the institutional change with the process of democratization of nations. Institutions and rules of democracies have a long tradition studies of political science since Aristotle and Machiavelli (Coccia, 2010). Modern approaches measure democracy with the quality of institutions and rules, such as the Freedom House Index of liberal democracy (for details, see Bogaards, 2007). In particular, the Freedom House Index of liberal democracy was launched by Raymond Gastil (1979) of the University of Washington in Seattle (USA). Gastil (1979) assigned ratings of political rights and civil liberties for 192 countries and 18 independent territories. The index of political rights consists of 10 criteria, which are grouped into three parts: electoral process, political pluralism and participation, and government functioning. This index ranges from 1 (best value) to 7, which is the worst value of democracy (cf., Munck and Verkuilen, 2002). Diamond (1986), Barro (1999), Coccia (2010) and Inglehart and Welzel (2005) apply this index for socioeconomic analyses.

This study focuses on *Freedom House (FH) Liberal Democracy standardized scale 100 pts,* 2000 year per country as well as on arithmetic mean of FH index from 1990-1996 (using data of countries from Norris, 2008a) to measure institutional change based on process of democratization. The year and time period of these variables are antecedents to response variables, given by innovative outputs, because the creation of institutions and institutional change generates effects on socioeconomic and technological factors in the medium-long run.

This study also considers other variables to assess institutions and institutional change of countries (cf., Kaufmann et al., 2008; Kaufmann et al, 1999; Norris, 2008a; Worldwide Governance Indicators, 2019; Thomas,

2010):

- *Kaufmann Voice and Accountability index in 2005* captures perceptions of the extent to which a country's citizens are able to participate in selecting their government, as well as freedom of expression, freedom of association, and a free media (Kaufmann et al., 1999, 2005, 2008, 2010; Worldwide Governance Indicators, 2019).
- *Kaufmann Political Stability and Absence of Violence/Terrorism 2005* measures perceptions of the likelihood of political instability and/or politically motivated violence, including terrorism (Worldwide Governance Indicators, 2019; cf., Coccia, 2018d)
- *Kaufmann government regulatory quality 2005* capturing perceptions of the ability of government to formulate and implement sound policies and regulations that permit and promote private sector development (Worldwide Governance Indicators, 2019)
- *Kaufmann Rule of Law 2005* capturing perceptions of the extent to which agents have confidence in and abide by the rules of society, and in particular quality of contract enforcement, property rights, police, and courts that also reduce the likelihood of crime and violence (Worldwide Governance Indicators, 2019; cf., Coccia, 2017e)
- Finally, *Kaufmann Control of Corruption 2005* capturing perceptions of the extent to which public power is exercised for private gain, including both petty and grand forms of corruption, as well as "capture" of the state by elites and private interests (Worldwide Governance Indicators, 2019)

– Innovative outputs, technology and examples of technological innovation

The second term of the relationship, analyzed here, is technology. It has numerous connotations, ranging from an object to a pool of applied scientific knowledge. Technology is based on inventions and innovations (Coccia, 2019a, b, c, d; Coccia and Watts, 2020). Invention is a commercially promising product or service based on new science or technology. Innovation is the successful entry of a new science or technology-based product or process into a particular market. The Pythagorean concept of technology focuses on patent statistics (Sahal, 1981). In this case, technological change is conceived in terms of the number of inventions patented. As a matter of fact, patterns of technological innovation can be measured with patents, which are an indicator of innovative outputs (Steil et al., 2002). In fact, economic literature gives particular attention to how innovators can appropriate returns by patents and intellectual property rights, which have an increasingly important role in the innovation and economic performance of countries. The increasing use of patents to protect inventions by private and public organizations is closely connected to recent evolutions in innovation processes that have become increasingly competitive, co-operative, global and more reliant on new entrants and technology–based firms (Coccia, 2019a, b, c). Cohen et al. (2001) demonstrate that patent protection is the central means for investors to reap returns in some sectors, such as pharmaceutical, fine chemical products, agricultural chemicals, etc. In fact, a patent protects the owner of the invention for a limited period of time, generally 20 years (Hall, 2007). In addition, Chen (2008) shows a significant positive effect of patent laws on invention rates. In short, a vast economic literature converges towards patents as measures of innovation (Jaffe and Trajtenberg, 2005). More specifically, the contribution here uses patent applications of residents to assess innovative potential of countries and overcome the distortion that patent applications to patent office can be also filed by residents in other countries. Patent applications filed by residents are applications filed with a national patent office for exclusive rights to inventions − a product or process that provides a new way of doing something or offers a new technical solution to a problem. However, patents as sources of innovation can have some limits: for instance, transaction costs and disclosure rules vary among countries. Considering this problem, the robustness of the analysis here based patent statistics is integrated with data of the adoption and diffusion of other vital technological innovations given by: internet users per 100 inhabitants 2007 year, personal computers per 1000 people 2005 year, cellular mobile telephone subscribers per 100 inhabitants 2005 year and average cellular mobile telephone subscribers per 100 inhabitants, 1995-2001 period, using data by Norris (2008a).

*1.3 Data analysis procedure*

*Firstly*, variables are analyzed with descriptive statistics based on mean, std. deviation, skewness and kurtosis to assess normality of distribution and, if necessary to fix distributions of variables with a *log*-transformation. Descriptive analysis and other statistical analyses of the sample under study are also done categorizing the countries with (cf., Norris, 2008a):

*a)* the type of democracy, given by: *Free* (higher level of democratization), *Partially Free* (average level of democratization) and *Not Free* (lower level of democratization).

*b)* the type of economy measured with the level of Gross Domestic Product per capita (GDPPC) in PPP

(Purchasing Power Parity) 2006 year (World Bank, 2009): i.e., countries with *High* ($15,000+), *Medium* ($2,000-$14,999) and *Low* ($2000 or less).

This analysis can show differences between countries on *how* institutional change, based on higher levels of democratization of nations, affects other variables of institutional change, wealth of nations, innovative outputs and adoption of new technologies.

*Secondly*, relationship between variables is analyzed considering a linear model of simple and multiple regression. The response variables of these models are innovative outputs and adoption of critical technological innovations (see previous sections). Explanatory variables are given by measures of institutional change and wealth of nations. Response variable has in general *a lag of 5 years* in comparison with explanatory variables to consider long-run effects on economic systems.

The operationalization of the model with simple regression analysis is specified as follows:

$$\log y_t = \alpha + \beta \log x_t + u_t \qquad [1]$$

$\alpha$ is a constant; $\log$ has base $e = 2.7182818$; $t$=time; $u_t$ = error term

$y_t$ (response variable) is Internet users per 100 inhabitants 2007y, Personal computers per 1000 people 2005y, Cellular mobile telephone subscribers per 100 inhabitants 2005y.

$x_t$ (explanatory variable) is a measure of the Freedom House (FH) Liberal Democracy standardized scale 100 pts 2000. In multiple regression analysis, the model also considers another explanatory variable given by GDP per capita PPP 2005y. *Note* that *y*=year.

Other models consider the following variables:

$y_t$ is a given by patents of residents per million people average1995-2001 or cellular mobile telephone subscribers per 100 inhabitants, average 1995-2001 period

$x_t$ is FH Liberal Democracy standardized scale 100 pts 1990-1996 and/or GDP per capita PPP average 1994-2000 period

The relationship [1] is analyzed using Ordinary Least Squares (OLS) method for estimating the unknown parameters in a linear regression model. Statistical analyses are performed with the Statistics Software SPSS® version 24.

**RESULTS**

Table 1  Parametric estimates of the relationship of institutional change leading to democratization on technological variables (*simple regression analysis*)

| RESPONSE VARIABLE | Constant $\alpha$ (St. Err.) | Coefficient $\beta$ (St. Err.) | $R^2$ adj. (St. Err. of the Estimate) | $F$ (sign.) |
|---|---|---|---|---|
| *log* Internet users per 100 inhabitants, 2007y | −3.47*** (0.79) | 1.44*** (0.19) | 0.23 (1.48) | 55.48 (0.001) |
| *log* Personal computers per 1000 people, 2005y | −0.48*** (1.48) | 1.10*** (0.37) | 0.13 (1.51) | 9.01 (0.004) |
| *log* Cellular mobile telephone subscribers per 100 inhabitants, 2005y | −1.81*** (0.68) | 1.23*** (0.17) | 0.25 (1.18) | 55.79 (0.001) |

Explanatory variable: *log* Freedom House Liberal Democracy standardized scale 100 pts, 2000y

*Note*: *** significant at 1‰; y=year

Table 1 shows the estimated relationship of technological variables on level of institutional change based on democratization. The regression coefficient β suggests that a 1% increase in the level of democratization increases:

− the expected Internet users by 1.44% (*p*-value < .001). $R^2$ value indicates that about 23% of the variation in Internet users can be attributed linearly to institutional change based on democratization

− the expected personal computer by 1.10% (*p*-value < .001). $R^2$ value indicates that about 13% of the variation in personal computer can be attributed linearly to institutional change based on democratization

− the expected cellular mobile by 1.23% (*p*-value < .001). $R^2$ value indicates that about 25% of the variation in cellular mobile can be attributed linearly to institutional change based on democratization

These relationships are illustrated in Figure 2-3-4

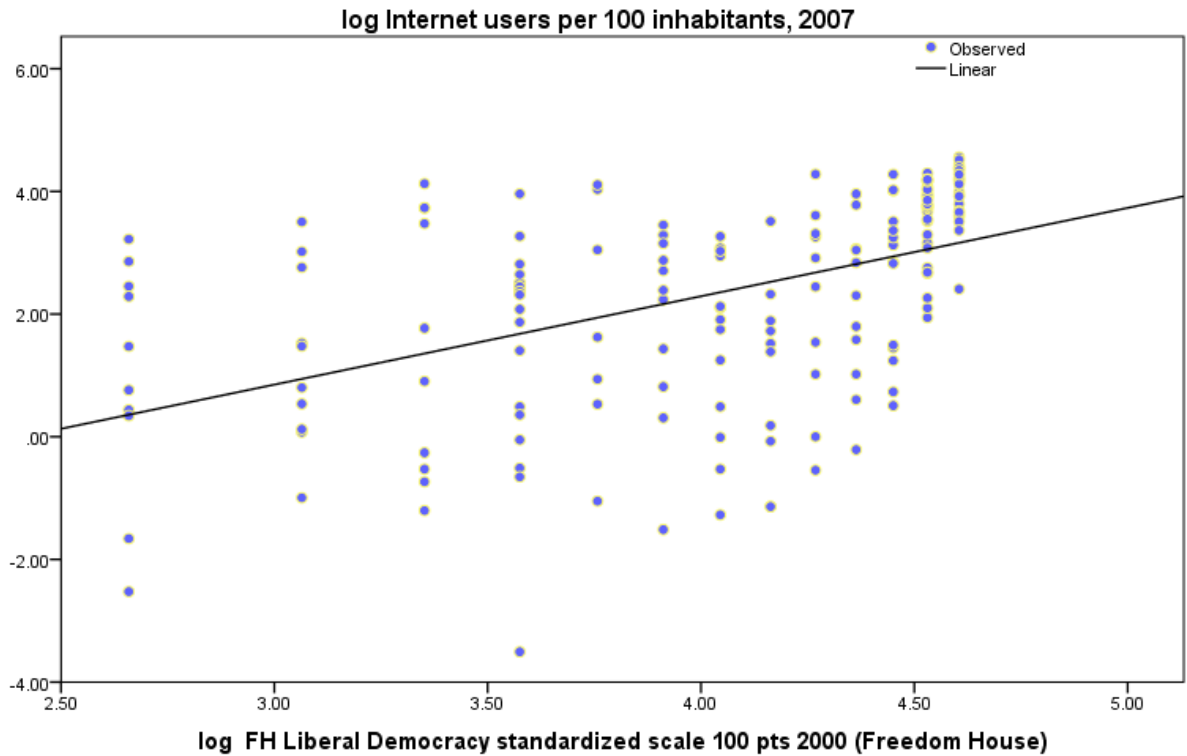

Figure 2 – Estimated relationship of institutional change, based on democratization, on internet users across countries (*log-log* scale)

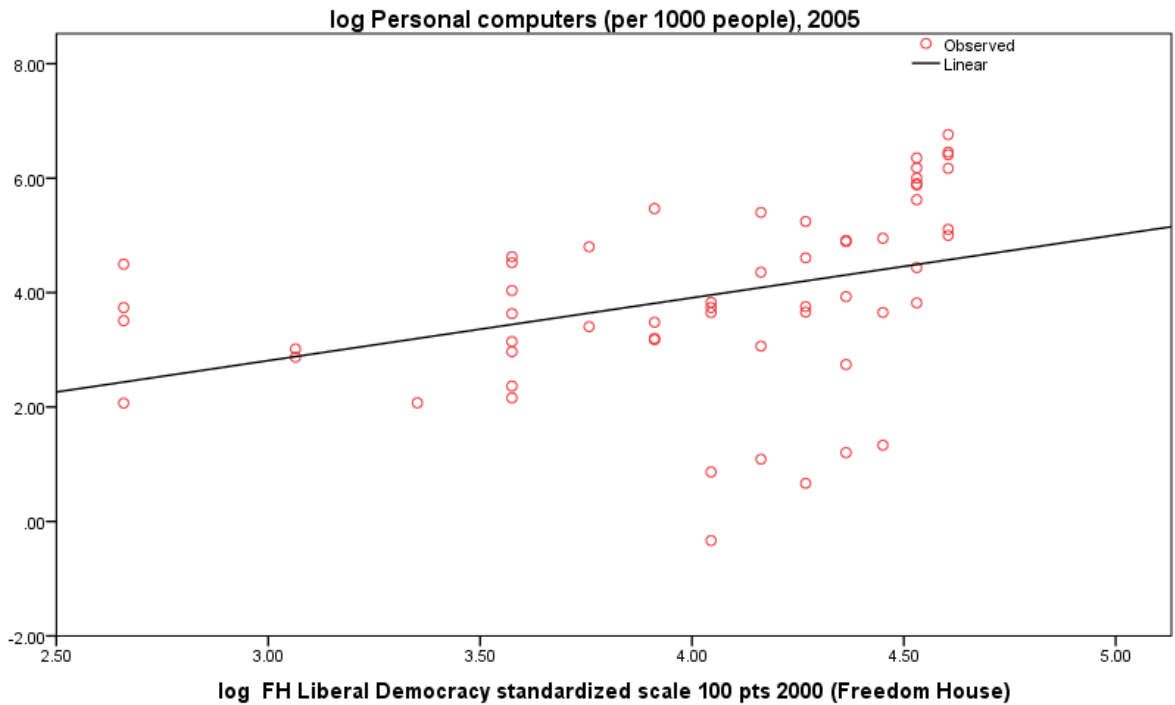

Figure 3 – Estimated relationship of institutional change, based on democratization, on personal computer across countries (*log-log* scale)

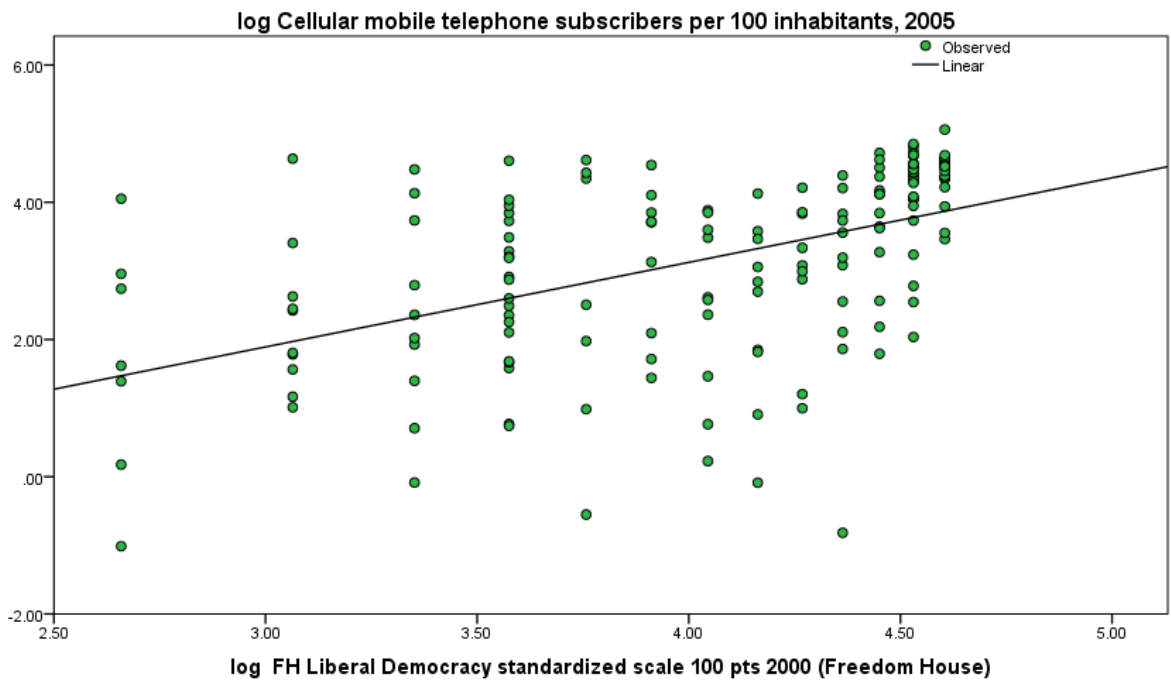

Figure 4 – Estimated relationship of institutional change, based on democratization, on cellular mobile telephone across countries (*log-log* scale)

Table 2  Parametric estimates of the relationship of institutional change, based on democratization, on technological variables (*multiple regression analysis*)

Explanatory variables:
*log* Freedom House Liberal Democracy standardized scale 100 pts, 2000y
*log* GDP per capita PPP 2005y

| RESPONSE VARIABLE | Constant α (St. Err.) | *log* FH Liberal Democracy standardized scale 100 pts 2000 Coefficient $\beta_1$ (St. Err.) | *log* GDP per capita PPP 2005 Coefficient $\beta_2$ (St. Err.) | $R^2$ adj. (St. Err. of the Estimate) | *F* (sign.) |
|---|---|---|---|---|---|
| *log* Internet users per 100 inhabitants, 2007y | –4.65*** (0.58) | 0.19 (0.16) | 0.81*** (0.05) | 0.67 (0.93) | 172.71 (0.001) |
| *log* Personal computers per 1000 people, 2005y | –1.86*** (0.96) | –0.26 (0.27) | 0.91*** (0.08) | 0.73 (0.85) | 72.83 (0.001) |
| *log* Cellular mobile telephone subscribers per 100 inhabitants, 2005y | –2.60*** (0.45) | 0.20 (0.12) | 0.65*** (0.04) | 0.72 (0.69) | 196.74 (0.001) |

Note: *** significant at 1‰; y=year

Table 2 shows the estimated relationship, with multiple regression analysis, of technological variables on level of democratization and GDP per capita across countries. The first partial regression coefficient shows that the effect of democratization is not significant, whereas the second coefficient of partial regression shows that a 1% increase in the level of GDP per capita, fixed the level of democratization, increases:

− the expected Internet users by 0.81% (*p*-value < .001). $R^2$ value indicates that about 67% of the variation in Internet users can be attributed linearly to institutional change of democratization and GDP per capita

− the expected personal computer by 0.91% (*p*-value < .001). $R^2$ value indicates that about 73% of the variation in personal computer can be attributed linearly to institutional change of democratization and GDP per capita

− the expected cellular mobile by 0.65% (*p*-value < .001). $R^2$ value indicates that about 72% of the variation in cellular mobile can be attributed linearly to institutional change of democratization and GDP per capita

Table 3 shows that institutions and institutional change in *free countrie*s–with a higher level of democratization–rather than partly and not free countries–with a lower level of democratization–, have a higher GDP per capita, adoption and diffusion of technologies under study. These results are underpinned with better governance indicators given by higher stability, higher regulatory quality, rule of law and control of corruption. Figure 5 shows the level of variables considering the categorization of countries in Free (higher level of democratization), Partially Free (average level of democratization) and Not Free (lower level of democratization). Results confirm that countries with institutions and institutional change based on higher levels of democratization provide better indicators of governance, emergence, adoption and diffusion of innovation (cf., Coccia, 1999, 2004, 2006, 2008a, 2018e; Coccia and Wang, 2015). The logical sequence of these findings are in figure 6.

Table 3    Descriptive statistics based on different levels of democracy

|  | Countries Free | | Partly Free | | Not Free | |
|---|---|---|---|---|---|---|
|  | Mean | SD | Mean | SD | Mean | SD |
| FH Liberal Democracy standardized scale 100 pts 2000 | 90.33 | 10.10 | 53.55 | 15.26 | 26.73 | 9.10 |
| Kaufmann voice and accountability 2005 | 0.85 | 0.55 | -0.48 | 0.41 | -1.33 | 0.43 |
| Kaufmann political stability 2005 | 0.64 | 0.62 | -0.65 | 0.76 | -0.68 | 1.00 |
| Kaufmann government effectiveness 2005 | 0.63 | 0.86 | -0.57 | 0.65 | -0.80 | 0.71 |
| Kaufmann government regulatory quality 2005 | 0.65 | 0.76 | -0.48 | 0.61 | -0.91 | 0.82 |
| Kaufmann rule of law 2005 | 0.64 | 0.80 | -0.59 | 0.66 | -0.81 | 0.75 |
| Kaufmann corruption 2005 | 0.62 | 0.91 | -0.57 | 0.62 | -0.72 | 0.69 |
| GDP per capita annual growth rate (%) 1975-2002 | 1.59 | 1.99 | -0.08 | 2.65 | 0.51 | 4.40 |
| GDP per capita annual growth rate (%) 1990-2002 | 1.89 | 1.73 | 0.78 | 3.53 | 1.73 | 4.46 |
| GDP per capita PPP 2005 | $11,329.38 | $12,030.65 | $2,252.44 | $4,660.43 | $3,050.43 | $6,055.47 |
| Internet users per 100 inhabitants 2007 | 40.54 | 25.10 | 11.01 | 13.58 | 11.74 | 14.81 |
| Personal computers (per 1000 people) 2005 | 246.95 | 243.26 | 60.25 | 74.30 | 43.17 | 36.67 |
| Cellular mobile telephone subscribers per 100 inhabitants 2005 | 66.19 | 36.02 | 25.69 | 27.22 | 23.82 | 26.47 |

*Note*: SD= Standard deviation

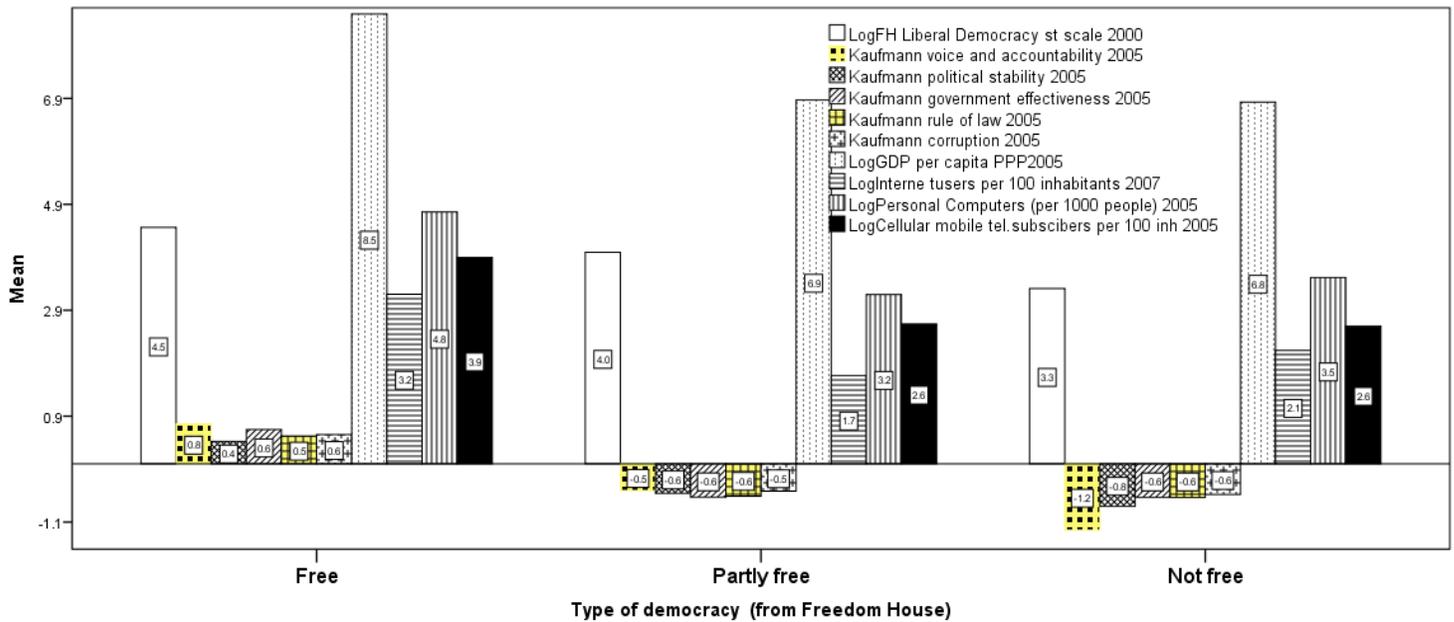

Figure 5 – Clustered bars of key variables per type of democracy. Note that some variables are in *log* scale to improve the visual representation of bar graphs

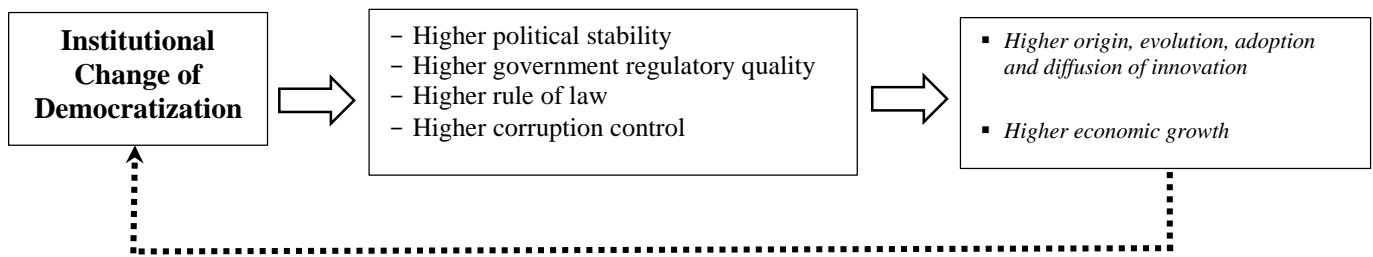

Figure 6 – Relation running from institutional change to patterns of technological innovation, with positive feedbacks

Table 3 shows a high association between level of democratization and GDP per capita across countries (variability of data measured with standard deviation is high within sets, suggesting a high heterogeneity of countries). Table 4, using the categorization per type of economy, considers arithmetic mean of some new variables across countries, specifically: average FH Liberal Democracy standardized scale 100 pts 1990-1996 period, average GDP per capita PPP 1994-2000 period, average Patents of residents per million people 1995-2001 period, average Cellular mobile telephone subscribers per 100 inhabitants, 1995-2001 period. Results confirm that richer countries having high GDP per capita ($15,000+) and a higher level of democratization,

rather than poorer countries with a lower level of democratization, have a higher production of innovative outputs (measured with average patents per million people) and a higher adoption and diffusion of new technology of cellular mobile telephone over time.

Table 4   Descriptive statistics per type of economy, using GDP per capita in PPP

| | Countries | | | | | |
| --- | --- | --- | --- | --- | --- | --- |
| | High ($15,000+) | | Medium ($2,000-14,999) | | Low ($2000 or less) | |
| | Mean | SD | Mean | SD | Mean | SD |
| FH Liberal Democracy standardized scale 100 pts, 1990-1996 | 95.96 | 8.98 | 64.52 | 23.35 | 44.89 | 21.58 |
| GDP per capita PPP, 1994-2000 | $23,484.76 | $5,728.91 | $6,559.06 | $3,325.41 | $1,256.77 | $422.82 |
| Patents of residents per million people, 1995-2001 | 498.69 | 563.90 | 31.23 | 37.99 | 18.47 | 24.74 |
| Cellular mobile telephone subscribers per 100 inhabitants, 1995-2001 | 393.37 | 242.76 | 69.43 | 101.70 | 6.53 | 14.15 |

*Note*: SD= Standard deviation

Table 5 also shows the estimated relationship of technological variables on level of institution change measured with democratization across countries, using variables analyzed in table 4. The regression coefficient suggests that a 1% increase in the level of democratization increases:

− the expected average patents of residents per million people by 2.42% (*p*-value < .001). $R^2$ value indicates that about 27% of the variation in patents can be attributed linearly to democratization

− the expected average cellular mobile telephone subscribers per 100 inhabitants by 2.74% (*p*-value <.001). $R^2$ value indicates that about 37% of the variation in cellular mobile subscribers users can be attributed linearly to democratization

These relationships are illustrated in Figure 7 and 8.

Table 5   Parametric estimates of the relationship of institutional change, based on democratization, on technological variables (*simple regression analysis*)

| RESPONSE VARIABLE | Constant α (St. Err.) | Explanatory variable: *log* average FH Liberal Democracy standardized scale 100 pts 1990-1996 period Coefficient β (St. Err.) | $R^2$ adj. (St. Err. of the Estimate) | F (sign.) |
|---|---|---|---|---|
| *log* average Patents of residents per million people, 1995-2001 period | −6.87*** (0.77) | 2.42*** (0.18) | 0.27 (2.15) | 176.31 (0.001) |
| *log* average Cellular mobile telephone subscribers per 100 inhabitants, 1995-2001 period | −7.93*** (0.68) | 2.74*** (0.16) | 0.37 (1.91) | 284.87 (0.001) |

*Note*: *** significant at 1‰

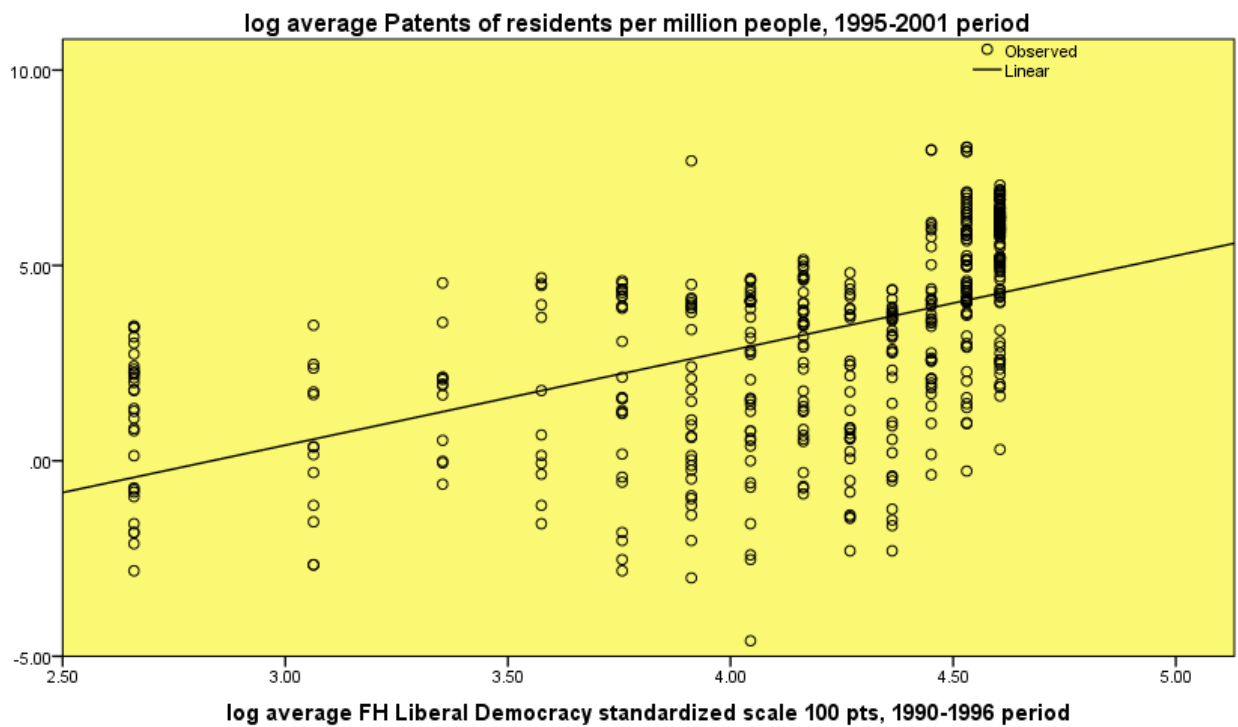

Figure 7 – Estimated relationship of institutional change, based on democratization, on patents per residents across countries (*log-log* scale)

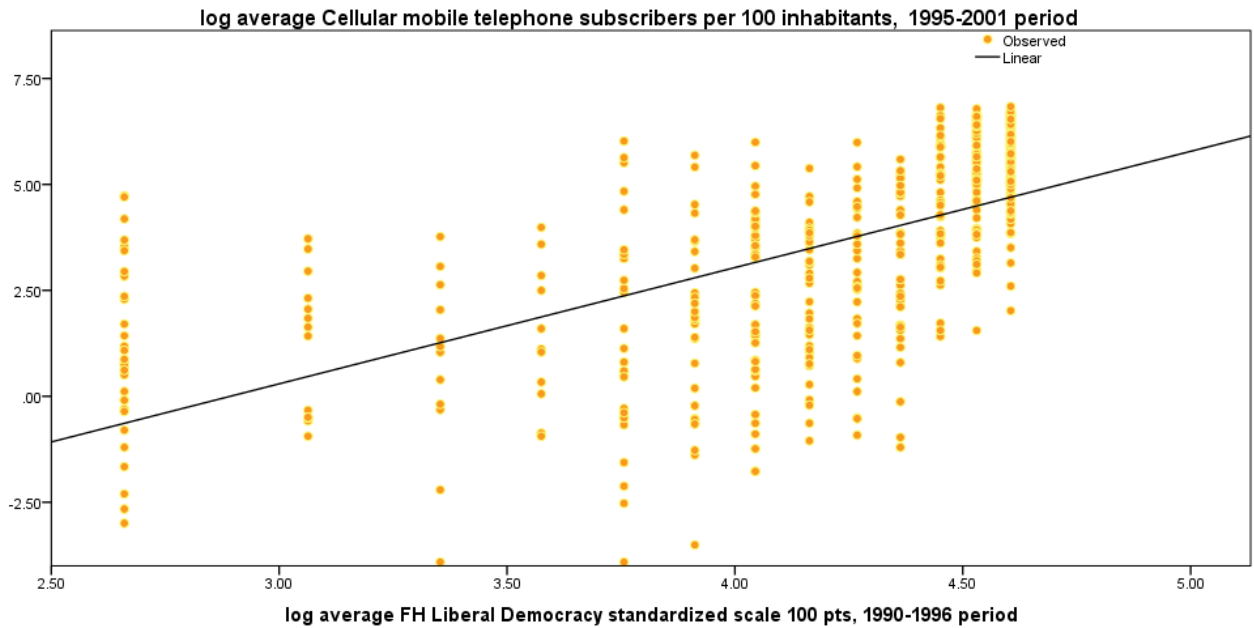

Figure 8 – Estimated relationship of institutional change, based on democratization, on cellular mobile telephone across countries (*log-log* scale)

The estimated relationship with multiple regression analysis of technological variables on level of democratization and GDP per capita across countries suggests similar results (Table 6).

As far as average patents of residents per million people (1995-2001 period) as response variable, the first partial regression coefficient shows that a 1% increase in the level of democratization, fixed the level of GDP per capita, increases:

− the expected average patents of residents per million people by 0.42% (*p*-value < .05)

The second partial regression coefficient shows that a 1% increase in the level of GDP per capita, fixed the level of democratization, increases:

− the expected average patents of residents per million people by 1.54% (*p*-value < .001)

$R^2$ value indicates that about 53% of the variation in patents can be attributed linearly to democratization and GDP per capita.

As far as average cellular mobile telephone subscribers per 100 inhabitants (1995-2001 period) as response variable, multiple regression analysis shows that (Table 6):

a 1% increase in the level of democratization, fixed the level of GDP per capita, increases:

− the expected average cellular mobile telephone subscribers per 100 inhabitants by 0.54% ($p$-value<.001)

whereas, a 1% increase in the level of GDP per capita, fixed the level of democratization, increases:

− the expected average cellular mobile telephone subscribers per 100 inhabitants by 1.69% ($p$-value<.001)

$R^2$ value indicates that about 71% of the variation in cellular mobile telephone subscribers can be attributed linearly to democratization and GDP per capita.

Table 6  Parametric estimates of the relationship of institutional change, based on democratization, on technological variables (*multiple regression analysis*)

| | | Explanatory variables: log Freedom House Liberal Democracy standardized scale 100 pts, 1990-1996 log GDP per capita PPP, 1994-2000 | | | |
|---|---|---|---|---|---|
| RESPONSE VARIABLE | Constant $\alpha$ (St. Err.) | log average FH Liberal Democracy standardized scale 100 pts 1990-1996 Coefficient $\beta_1$ (St. Err.) | log average GDP per capita PPP 1994-2000 Coefficient $\beta_2$ (St. Err.) | $R^2$ adj. (St. Err. of the Estimate) | $F$ (sign.) |
| *log* Patents of residents per million people average 1995-2001 | −12.13*** (0.69) | 0.42** (0.19) | 1.54*** (0.09) | 0.52 (1.72) | 270.61 (0.001) |
| *log* Cellular mobile telephone subscribers per 100 inhabitants average 1995-2001 | −13.69*** (0.52) | 0.54*** (0.14) | 1.69*** (0.07) | 0.71 (1.30) | 591.98 (0.001) |

*Note*: *** significant at 1‰; ** significant at 5%

Finally, table 7 shows the estimated relationships with multiple regression analysis of technological variables on level of democratization and GDP per capita across countries, considering the type of economy based on three categories of GDP per capita PPP, 2006 year: i.e., *High* $15,000+, *Medium* $2,000-$14,999, *Low* $2,000 or less. Because of high correlation between level of democratization and GDP per capita across countries, the categorization in table 7 provides similar results to the categorization of countries in *Free* (higher level of democratization), *Partially Free* (average level of democratization) and *Not Free* (lower level of democratization).

As far as average patents of residents per million people (1995-2001 period) as response variable, the first partial regression coefficient is not significant, whereas the second one shows that a 1% increase in the level of GDP per capita, fixed the level of democratization, increases the expected average innovative outputs mainly in poor and richer countries (by 2.45%, *p*-value < .001; by 2.43%, *p*-value < .001, respectively), rather than countries with a medium income per capita. $R^2$ value of three models has a range between 15-19%.

As far as average cellular mobile telephone subscribers per 100 inhabitants (1995-2001 period) as response variable, multiple regression analysis shows the following results (Table 7):

the first partial regression coefficient shows that a 1% increase in the level of democratization, fixed the level of GDP per capita, increases:

− the expected average cellular mobile telephone subscribers per 100 inhabitants mainly in poor countries by 0.64% (*p*-value<.05), whereas in countries with medium income per capita by 0.42% (*p*-value<.05). In rich countries the coefficient is not significant.

The second partial regression coefficient shows that a 1% increase in the level of GDP per capita, fixed the level of democratization, increases:

- the expected average cellular mobile telephone subscribers per 100 inhabitants mainly in countries with a medium income per capita by 2.1% (*p*-value<.001), after poor countries by 1.34% (*p*-value<.001) and finally rich countries by 1.09% (*p*-value<.001)

$R^2$ value is rather low except the estimated relation of countries with medium income per capita where about 43% of the variation in cellular mobile telephone subscribers can be attributed linearly to democratization and GDP per capita. The lower effect of institutional change and economic growth on cellular mobile telephone technology in developing countries, it can be due to low development of system of information and communication networks, of its use and low technical improvements over time. Instead, in rich countries the lower impact can be likely explained with decreasing return effects of the development of information and communication networks.

Table 7  Parametric estimates of the relationship of institutional change on technological variables per type of economy (*multiple regression analysis*)

*Explanatory variables*:
*log* Freedom House Liberal Democracy standardized scale 100 pts, 1990-1996
*log* GDP per capita PPP, 1994-2000

| RESPONSE VARIABLE | Constant α (St. Err.) | *log* average FH Liberal Democracy standardized scale 100 pts 1990-1996 Coefficient β₁ (St. Err.) | *log* average GDP per capita PPP 1994-2000 Coefficient β₂ (St. Err.) | $R^2$ adj. (St. Err. of the Estimate) | F (sign.) |
|---|---|---|---|---|---|
| *log* average Patents of residents per million people average 1995-2001 period | | | | | |
| Countries with *low* income per capita $2000 or less | −18.78*** (4.75) | 0.67 (0.43) | 2.45*** (0.65) | 0.17 (2.24) | 8.79 (0.001) |
| Countries with *medium* income per capita $2,000-$14,999 | −9.52*** (1.78) | 0.25 (0.24) | 1.25*** (0.23) | 0.15 (1.68) | 22.56 (0.001) |
| Countries with *high* income per capita $15,000+ | −17.04*** (4.75) | −0.37 (0.71) | 2.43*** (0.40) | 0.19 (1.06) | 8.79 (0.001) |
| *log* average Cellular mobile telephone subscribers per 100 inhabitants, 1995-2001 period | | | | | |
| Countries with *low* income per capita $2000 or less | −11.51** (3.74) | 0.64* (0.34) | 1.34** (0.51) | 0.10 (1.76) | 5.59 (0.005) |
| Countries with *medium* income per capita $2,000-$14,999 | −16.79*** (1.45) | 0.42* (0.20) | 2.10*** (0.19) | 0.43 (1.37) | 94.72 (0.001) |
| Countries with *high* income per capita $15,000+ | −3.13 (3.30) | −0.47 (0.52) | 1.09*** (0.29) | 0.07 (0.78) | 6.97 (0.001) |

*Note*: *** significant at 1‰; ** significant at 1%, * significant at 5%

## DISCUSSION

Considering the results just mentioned, the fundamental question is:

*How does institutional change, based on democratization, support patterns of technological innovation?*

Zuazu (2019) argues that the interplay between democracy and technological development is crucial to the economic performance of industries. He shows a technologically-conditioned effect of democracy. In particular, political system changes towards democracy are growth-enhancing for industries close to the World Technology Frontier (WTF) but may have a negative effect on backward industries. In this context, a vital role is played by linkages between democracy, economic freedom and regulation (De Haan and Sturm 2000, 2003; Lundstrom 2005; Djankov et al. 2002; Rode and Gwartney 2012). Aghion et al. (2009) show theoretically and empirically that democracy promotes innovation in advanced industries. Moreover, freedom of entry is also a determinant for sectors close to the WTF since, as suggested by Aghion et al. (2008), entry of new firms and competition spur innovation towards high levels of technological development but discourage innovation in backward sectors. Coccia (2010) shows that democratization is a driving force for technological change: most free countries, measured with liberal, participatory, and constitutional democracy indices, have a higher level of technology than less free and more autocratic countries. In fact, *democracy richness* generates a higher rate of technological innovation with fruitful effects for the wellbeing and wealth of nations (cf., Bell and Staeheli, 2001). In general, a fruitful relation between technology, economic growth, institutional change and democracy can be supported by three factors:

 a) economic freedom,

b) regulation and

c) economic and political stability, good economic governance and higher level of education system.

a) The relation between democracy and economic freedom

Studies suggest that democracy is conducive to economic freedom (Pitlik and Wirth, 2003; Pitlik,2008). De Haan and Sturm (2003) show that the increase in economic freedom between 1975 and 1990 in developing countries was driven by the level of political freedom. Rode and Gwartney (2012) confirm these results using a panel data set covering 48 political transitions from authoritarianism to democracy since the mid-1970s. An overall, positive association of economic freedom with economic growth is also suggested by Doucouliagos and Ulubasoglu (2006). In general, studies seem to show that institutional change of democracy fosters economic growth and new technological pathways through its effect on economic freedom and regulation (Zuazu, 2019).

b) The relation between democracy and regulation

Democracy shapes the intervention of the state in the economy and determines the level and quality of regulation. Djankov et al. (2002, 2006) and Jalilian et al. (2007) show that more democratic countries and limited intervention of governments have lighter regulation and thus lower market-entry barriers (cf., Weyland, 2002). In short, democratization can provide higher levels of political accountability that reduce protection of vested interests, so that the resulting lower market-entry barriers work in turn in favor of sectors that are better able to adapt to new economic scenarios and pathways of technological change.

c) the relation between democracy, political stability, economic governance and higher level of education system

Democracy is associated with more stable political systems that provide benefits for higher education systems, institutions and paths of technological and economic change (cf., Alesina and Perotti,1996, Rodrik, 2000; Rodrik and Wacziarg, 2005). Taverdi et al. (2019) show that the effectiveness of governance increases with economic development and education of nation (cf., Farazmand and Pinkowski, 2006; Farazmand, 2019). In fact, political and economic stability and the securing of property rights make democracies more appropriate environments for technological innovation than oligarchies (Acemoglu, 2008; cf., Coccia, 2016a, 2017d). Milner (2006) provides evidence on the crucial role of regime type in the diffusion of Internet. Gao et al. (2017) argue that democracy is positively associated with innovation in an indirect way. Zuazu (2019) claims that industries with a comparative advantage in new technologies are more likely to grow in democratic countries, since democracies are political systems associated with higher levels of economic freedom, investment in higher education systems and lower limits on market entry. By contrast, new investment opportunities are reduced when market-entry barriers are high, property rights are not properly enforced and nations have political and economic instability. Finally, Dixit (2009) states that economic governance is the structure and functioning of the legal and social institutions that support economic activity and economic transactions by protecting property rights, enforcing contracts, and taking collective action to provide physical and organizational infrastructure. Overall, then, markets, economic activity and transactions function well in the presence of a good economic governance based on institutional change directed to democratization of countries. Table 3 shows a good synthesis of these findings for advanced and emerging economies.

## CONCLUSION AND LIMITATIONS

Technological and institutional change cannot be discussed in isolation from each other. This interaction can explain economic growth and social change as well as wealth and wellbeing of nations (Kaiserfeld, 2015). In general, differences in institutional arrangements between countries can explain why new technological path creation takes place more easily in some regions than others. Evidence of the impact of institutional differences across nations has been provided with respect to economic policy within different varieties of capitalism by Hall and Soskice (2001; cf., Coccia, 2017), and with respect to national systems of innovation by Lundvall (1995) and Freeman and Soete (1997). At the local level, Gertler (2010) argues that different institutions contribute to different pathways of economic development in different regional settings. Chlebna and Simmie (2018) show that successful invention, innovation and diffusion of new technologies require the co-evolution of vital institutions.

This contribution here shows a main insight: institutional change based on democratization is a determinant of technological and economic change, i.e. initially, democratization creates institutions and institutional change that are preconditions (factors that set the stage over the long run) to support paths of technological innovation and, as a consequence, of economic growth of nations (cf., Grossman and Helpman, 1991). *Subsequently*, the relation between institutional change and technological development is intertwined over time. In short, institutional change leading to higher level of democratization generates economic freedom, a better higher education system and economic governance supporting a greater production and adoption of technology for technical and economic change of countries. These results are important, very important in the modern era to sustain technology and economic growth in view of the accelerating globalization and expansion of markets (cf., Coccia, 2018f, 2019g, 2019i).

In particular, countries to achieve, sustain and improve democratization need bring out the value of people and to increase the education of human capital and, as a consequence, the accumulation of intangible capital based on knowledge that has a greater and greater influence on technology production, diffusion and on the competitive advantage of countries (Coccia, 2004, 2008a, 2009, 2018a, 2019e). Democracy has some drawbacks that may generate political and economic crisis, as showed in the course of economic history, but democratic institutions have several advantages in comparison to other political systems because they support period of peace and economic stability ("Democratic Peace") associated with technological progress, economic growth and wellbeing of nations (Coccia, 2019d, p. 5). Modelski and Perry III (2002) argue that the main advantage of democracy lies in its capacity to enhance cooperation and manage conflict (cf., Coccia, 2019f). People increasingly prefer to live in democracies that are contagious and continuously spreading. Therefore, sustainable institutional change within democratic settings should be much more diffused across emerging market economies and improved where already applied (i.e., developed countries with consolidated democracy). However, the causal effect of democratization on technological and economic change needs to be further investigated considering several historical, social, economic and institutional factors that can affect this complex relationship. The findings of this chapter lead to the conclusion that policy makers need to be cognizant that institutional change based on democratic pathways sustains economic stability and a high quality of higher education system, which are main preconditions for the origin, diffusion and utilization of technology and economic growth within and between economic systems (cf., Coccia 2005, 2005a, 2006, 2008, 2016a, 2017d). Hence, political economy of growth should be designed considering the joint coevolution of democratic and social systems in order to support a fruitful institutional change and good economic governance for technical change directed to distribute total wealth among the widest fraction of population (cf., Bellah et al., 1991; Dixit, 2009; Farazmand and Pinkowski, 2006; Farazmand, 2019; Selznick, 1992; Wolfe, 1989). Moreover, technological revolution generates a disequilibrium between a socio-institutional framework geared to supporting the deployment of the old paradigm and the new techno-economic sphere brimming with change (Aglietta, 1976; Perez, 2004). Thus, long wave transitions are processes of creative destruction supporting economic, social and institutional change in advanced and emerging countries. These insights are important, very important for economists, policy makers and politicians, since they can propose best practices of institutional change supporting a higher democratization that, as proven, can foster technological progress, economic growth of countries, and therefore the wealth and wellbeing of nations (cf., Coccia, 2010).

To conclude, the challenge for institutional scholars and economists of technology is to continue the theoretical and empirical exploration of this *terra incognita* of the relation of institutions and institutional change with pathways of technological innovation considering more and more interdisciplinary approaches to exploit the diversity of viewpoints that generate scientific breakthroughs and appropriate socio-institutional policies to improve human interactions directed to support a fruitful technological and economic development in society.

33 | P a g e